\documentstyle[aps,twocolumn]{revtex}
\newcommand{\vect}[1]{\mbox{\boldmath $#1$}}
\input{psfig}

\font\fvmib=cmmib10 scaled 500
\font\sixmib=cmmib10 scaled 600

\font\eitmib=cmmib10 scaled 800
\font\tenmib=cmmib10
\font\twlvmib=cmmib10 scaled \magstep1

\newfam\mibfam \def\mib{\fam\mibfam\tenmib}
\textfont\mibfam=\tenmib \scriptfont\mibfam=\tenmib
\scriptscriptfont\mibfam=\fvmib

\newfam\Mibfam 
\textfont\Mibfam=\twlvmib \scriptfont\Mibfam=\eitmib
\scriptscriptfont\Mibfam=\sixmib

\begin{document}
\draft


\title{Spin-triplet superconductivity due to antiferromagnetic 
spin-fluctuation in Sr$_2$RuO$_4$}

\author{Takeshi Kuwabara\thanks{e-mail address: t-kuwabara@bea.hi-ho.ne.jp} 
and 
Masao Ogata\thanks{Address from April 2000, Dept.\ of Physics, Univ.\ of Tokyo,
Hongo, Bunkyo-ku 113-0033 Tokyo}}

\address{Department of Basic Science, Graduate School of Arts and Sciences,
University of Tokyo, Komaba 3-8-1, Tokyo 153-8902, Japan} 

\date{\today}

\maketitle

\begin{abstract}
A mechanism leading to the spin-triplet superconductivity is proposed 
based on the antiferromagnetic spin fluctuation.  
The effects of anisotropy in spin fluctuation on the Cooper pairing 
and on the direction of $\mib{d}$ vector are examined in the one-band 
Hubbard model with RPA approximation.   
The gap equations for the anisotropic case are derived and 
applied to Sr$_2$RuO$_4$.  
It is found that a nesting property of the Fermi surface together with 
the anisotropy leads to the triplet superconductivity with the 
${\mib{d}}=\hat{\mib{z}}(\sin{k_x}\pm i\sin{k_y})$, 
which is consistent with experiments.  
\end{abstract}


Since the discovery of superconducting phase in Sr$_2$RuO$_4$\cite{maeno1}, 
much effort has been paid for understanding its exotic properties.  
Among several interesting natures, the most fascinating one is that 
it is a spin-triplet superconductor confirmed by NMR experiment\cite{ishida1}. 
While most superconductors found during several decades are singlet, 
the only exceptions were $^3$He and UPt$_3$.  
Therefore the fact that the triplet pairing is realized in 
Sr$_2$RuO$_4$ has attracted much attention.  
While UPt$_3$, the second example of spin-triplet superconductor, 
has a complicated electronic structure, Sr$_2$RuO$_4$ has a rather simple 
electronic state\cite{maeno1}.  
Thus clarifying the microscopic mechanism of superconductivity 
in Sr$_2$RuO$_4$ is very important for understanding 
the triplet superconductors in general.  

In $^3$He, Cooper pairs are formed due to ferromagnetic spin fluctuations 
peaked at $\vect{q}=\vect{0}$\cite{Scalapino,nakajima}.   
Therefore it is natural to expect the origin of the triplet pairing in 
Sr$_2$RuO$_4$ is also ferromagnetic spin 
fluctuation\cite{mazin1,Sigrist2}.
This assumption has been believed to be justified by NMR 
experiments\cite{Imai,Mukuda1,Mukuda2}.   
However the recent neutron scattering experiment has shown that 
there exists a significant peak near $\vect{q}_0=(\pm 2\pi/3,\pm 2\pi/3)$ 
and no sizable ferromagnetic spin fluctuation\cite{Sidis}.  
Thus it is difficult to assume that the spin fluctuation 
near $\vect{q}_0$ plays no role in the Cooper pairing in Sr$_2$RuO$_4$. 
(In the following discussion we call this fluctuation as 
antiferromagnetic (AF) spin fluctuation, for simplicity.)
However this AF fluctuation leads to the singlet superconductivity rather 
than the triplet superconductivity as expected in analogy to 
high-$T_c$ cuprates\cite{mazin1}.

In this paper we propose a mechanism which gives the triplet 
pairing even if the spin fluctuation is AF.  
We find that the characteristic features of Sr$_2$RuO$_4$ are twofold:
One is the anisotropy of the spin fluctuation found in NMR 
experiments\cite{Mukuda1,Mukuda2}, and 
the other is a nesting property with momentum $\vect{q}_0$ 
of the two-dimensional Fermi surface.  
We show that these two features explain the pairing in Sr$_2$RuO$_4$.

In addition to the competition between singlet and triplet pairing, 
the direction of the $\vect{d}$ vector, which is the order parameter 
of triplet superconductivity, is another interesting problem.  
We show that the anisotropy of the spin fluctuation also explains  
the experimental fact that 
the $\vect{d}$ vector is parallel to the $z$-direction\cite{ishida1}. 
First we extend the RPA formulation to the case of anisotropic spin 
fluctuation.  
Using the obtained effective interactions, we investigate the most stable 
pairing based on the weak-coupling gap equations.  
When the spin fluctuation is isotropic, the so-called d$_{x^2-y^2}$-wave 
pairing is the most stable.  
However when the anisotropy is increased, the state corresponding to 
$\hat{\vect{z}}(\sin{k_x}\pm i\sin{k_y})$, which is the prime candidate 
of Sr$_2$RuO$_4$, becomes the most stable.  

For the $\gamma$ band which is one of the three bands 
in Sr$_2$RuO$_4$\cite{oguchi}, we assume a two-dimensional 
effective Hamiltonian
\begin{equation}
H = H_0 + \frac{I}{2N}\sum_{kk'q\sigma} 
    c^{\dagger}_{k\sigma}c^{\dagger}_{k'-\sigma}c_{k'-q-\sigma}c_{k+q\sigma},
\label{Hubbard Hamiltonian}
\end{equation}
where $c_{k\sigma}$ is the annihilation operator of an electron with 
momentum $\vect{k}$ and spin $\sigma$.
We consider only the on-site Coulomb repulsion, $I$, as in the previous 
studies of spin-fluctuation mechanism.  
Among the three bands, we consider the $\gamma$ band consisting of 
the antibonding band of Ru 4d$_{xy}$ and O 2p$_\pi$ orbitals in this paper, 
because it has the largest density of states at Fermi energy 
and the superconductivity is considered to be realized predominantly 
in the $\gamma$ band\cite{Agterberg}.  
Although the spin fluctuation near $\vect{q}_0$ is understood from 
the nesting effect of $\alpha$ and $\beta$ bands\cite{mazin1},
we assume that the wave-number dependence of spin fluctuation is 
common in the three bands due to some interactions, such as 
spin-orbit couplings and/or Hund couplings.
In the following calculation, we use the $\gamma$ band.  
However the same discussions can be also applied to the 
$\alpha$ and $\beta$ bands.  

The anisotropy of spin fluctuation observed experimentally is 
implicitly included in the two-body Hamiltonian, $H_0$.  
Our purpose is not to investigate the origin of anisotropy in details
but to examine the role of the anisotropy to Cooper pairing.  
Therefore we introduce a phenomenological parameter $\alpha$ by 
\begin{equation}
\chi_{(+-,\ 0)}({\vect q})=\alpha \chi_{(\uparrow \uparrow,\ 0)}({\vect q}),
\label{chiani}
\end{equation}
where $\chi_{(\uparrow \uparrow,\ 0)}({\vect q})$ 
($\chi_{(+-,\ 0)}({\vect q})$) is the 
unperturbed static susceptibility of $z$ axis ($xy$ plane), which 
originates from $H_0$.  
The parameter $\alpha$ represents the anisotropy of spin fluctuation 
and we take $\alpha\leq1$ since NMR experiments show that 
$\chi_{(xx)}<\chi_{(zz)}$\cite{Mukuda1,Mukuda2}.  

Using this one-band model, we discuss the effective interactions 
between Cooper pairs due to spin fluctuations.  
Summation of bubble and ladder diagrams (i.e., RPA approximation) gives
\begin{eqnarray}
H_{\rm int} =&& -\sum_{kk's}V_{{\rm b.o}}(\vect{k}-\vect{k}') 
   c^{\dagger}_{ks}c^{\dagger}_{-ks}c_{-k's}c_{k's} \nonumber\\
&& +\sum_{kk's}V_{{\rm b.e}}(\vect{k}-\vect{k}') 
   c^{\dagger}_{ks}c^{\dagger}_{-k-s}c_{-k'-s}c_{k's} \nonumber\\
&& -\sum_{kk's}V_{{\rm lad}}(\vect{k}-\vect{k}') 
   c^{\dagger}_{ks}c^{\dagger}_{-k-s}c_{-k's}c_{k'-s},
\label{Vre}
\end{eqnarray}
with
\begin{eqnarray}
V_{{\rm b.o}}(\vect{k}-\vect{k}') &=& \frac{I}{N}
\frac{(I/N)\chi_{(\uparrow \uparrow,\ 0)}(\vect{k}-\vect{k}')}
{1-(I/N)^2\chi^2_{(\uparrow \uparrow,\ 0)}(\vect{k}-\vect{k}')}, \nonumber\\
V_{{\rm b.e}}(\vect{k}-\vect{k}') &=& \frac{I}{N}
\frac{(I/N)^2\chi^2_{(\uparrow \uparrow,\ 0)}(\vect{k}-\vect{k}')}
{1-(I/N)^2\chi^2_{(\uparrow \uparrow,\ 0)}(\vect{k}-\vect{k}')}, \nonumber\\
V_{{\rm lad}}(\vect{k}-\vect{k}') &=& \frac{I}{N}
\frac{(I/N)\chi_{(+-,\ 0)}(\vect{k}-\vect{k}')}
{1-(I/N)\chi_{(+-,\ 0)}(\vect{k}-\vect{k}')}.
\label{Vbo}
\end{eqnarray}
Here $V_{\rm b.o}$ ($V_{\rm b.e}$) comes from the summation of 
diagrams with odd (even) number of bubbles, 
and $V_{\rm lad}$ from the ladder diagrams. 
It is apparent that $V_{\rm b.o}$ is between the electrons with equal 
spins while $V_{\rm b.e}$ and $V_{\rm lad}$ are between those 
with the opposite spins.  

It is straightforward to derive the gap equations in the anisotropic case, 
using the method developed by Leggett\cite{Leggett}.  
First we introduce the operators 
\begin{eqnarray} 
t_{k}^{(0)}&=&\sum_{ss'}(\sigma_2)_{ss'}c_{-ks}c_{ks'}\label{t0}, \\
t_{k}^{(a)}&=&\sum_{ss'}(\sigma_2\sigma_a)_{ss'}c_{-ks}c_{ks'},\ 
{\rm for}\ a=1,\ 2,\ 3,\label{ta}
\end{eqnarray}
where $\sigma_a (a=1,2,3)$ are Pauli matrices.  
The operator $t_{k}^{(0)}$ ($t_{k}^{(a)}$) 
corresponds to spin singlet (triplet)
Cooper pairs.  In terms of these operators, the effective interaction 
(\ref{Vre}) can be rewritten as
\begin{eqnarray}
H_{\rm int}&=& \frac{1}{4}\sum_{kk'}
V_{{\rm sin}}(\vect{k}-\vect{k'}) t_{k}^{(0)\dagger}t_{k'}^{(0)} \nonumber \\
& & +\frac{1}{4}\sum_{kk'} \sum_{a=1}^3 
V_{\rm tri}^{(a)}(\vect{k}-\vect{k'}) t_{k}^{(a)\dagger}t_{k'}^{(a)}, 
\end{eqnarray}
where
\begin{eqnarray}
V_{{\rm sin}}(\vect{k}-\vect{k}')&\equiv& 
2[ V_{{\rm b.e}}(\vect{k}-\vect{k}')+V_{{\rm lad}}(\vect{k}-\vect{k}')],
\nonumber\\
V_{\rm tri}^{(1)}(\vect{k}-\vect{k}')&=&V_{\rm tri}^{(2)}(\vect{k}-\vect{k}')
\equiv -2V_{{\rm b.o}}(\vect{k}-\vect{k}'),\nonumber\\
V_{\rm tri}^{(3)}(\vect{k}-\vect{k}')&\equiv&
2[ V_{{\rm b.e}}(\vect{k}-\vect{k}')-V_{{\rm lad}}(\vect{k}-\vect{k}')].
\label{Vsin}
\end{eqnarray}

Since Sr$_2$RuO$_4$ has a long coherence length in $ab$ plane, 
$\xi_{ab}\approx 660$ \AA\cite{Maeno3}, 
we use mean-field approximation to $H_{\rm int}$.  
We restrict the discussion to unitary states because it is 
unrealistic to assume non-unitary states in Sr$_2$RuO$_4$\cite{ishida2}.
Requiring that there is no coexistence of singlet and triplet pairs,
we obtain the gap equations 
\begin{eqnarray}
\Delta(\vect{k}) &=& -\sum_{k'}V_{{\rm sin}}(\vect{k}-\vect{k}')
\Delta(\vect{k}')\Theta(E_{{\rm sin}}(\vect{k}')), \nonumber\\
d^{(a)}(\vect{k}) &=& -\sum_{k'}V_{\rm tri}^{(a)}(\vect{k}-\vect{k}')
d^{(a)}(\vect{k}')\Theta(E_{{\rm tri}}(\vect{k}')),
\label{gap eq}
\end{eqnarray}
where 
$\Theta(E)\equiv\frac{1}{2E}\mbox{tanh} \frac{\beta E}{2}$, 
$E_{{\rm sin}}^2(\vect{k})=\xi^2_k+
\Delta(\vect{k})\Delta^{\ast}(\vect{k})$, and 
$E_{{\rm tri}}^2(\vect{k})=\xi^2_k+
\vect{d}(\vect{k})\cdot \vect{d}^{\ast}(\vect{k})$ with
$\xi_k = \varepsilon_k -\mu$.
The singlet and triplet order parameters are defined as 
$\Delta(\vect{k}) = -\frac{1}{2}\sum_{k'}V_{\rm sin}(\vect{k}-\vect{k}')
\langle t^{(0)}_{k'}\rangle$ 
and 
$d^{(a)}(\vect{k}) = -\frac{1}{2}\sum_{k'}V_{\rm tri}^{(a)}(\vect{k}-\vect{k}')
\langle t^{(a)}_{k'}\rangle$, 
respectively.  
Here $\vect{d}(\vect{k})$ is the so-called $\vect{d}$-vector for the 
triplet superconductivity.  

In the system with the rotational symmetry in spin space, 
$\chi_{(\uparrow \uparrow,\ 0)}({\vect q})=\chi_{(+-,\ 0)}({\vect q})$ 
is satisfied and 
thus the relation $V_{\rm b.o}+V_{\rm b.e}=V_{\rm lad}$ holds.  
In this case, it is easy to see 
$V_{\rm tri}^{(1)}(\vect{k}-\vect{k}')=V_{\rm tri}^{(2)}(\vect{k}-\vect{k}')
=V_{\rm tri}^{(3)}(\vect{k}-\vect{k}')$.  

On the other hand, the gap equation in Eq.\ (\ref{gap eq})
for the triplet pairing becomes dependent on the direction of the 
$\vect d$ vector in the anisotropic case.  
It means that {\it $\vect{d}$ vector 
has some preferred direction if the triplet pairs are formed by 
anisotropic spin fluctuations}.  
This is naturally understood because the $\vect{d}$ vector 
is orthogonal to the spin direction of triplet Cooper pairs\cite{Vollhardt}.  
For the present case with 
$\chi_{(+-,\ 0)}({\vect q}) < \chi_{(\uparrow \uparrow,\ 0)}({\vect q})$ 
(i.e., $\alpha < 1$) which is applied to the Sr$_2$RuO$_4$, 
we can see from Eq.\ (\ref{Vbo}) 
that $V_{{\rm lad}}(\vect{k}-\vect{k}')$ is suppressed 
and the effective interaction $V_{\rm tri}^{(3)}(\vect{k}-\vect{k}')$
approaches $V_{{\rm sin}}(\vect{k}-\vect{k}')$.  
Consequently the triplet superconductivity with $d^{(3)}(\vect{k})$
(i.e., $\vect{d}\parallel\hat{\vect{z}}$) can be stabilized 
even due to the AF spin fluctuations.  

In order to determine the symmetry of the superconducting order 
parameter, we have to take account of their sign change 
along the Fermi surface.  
For the high-$T_{\rm c}$ superconductors, the AF spin fluctuation 
with momentum $(\pi,\pi)$ stabilizes the singlet d$_{x^2-y^2}$-wave 
superconductivity.  
In that case, the singlet order parameters 
$\Delta(\vect{k}')$ with $\vect{k}'=(\pi,0)$ and 
$\Delta(\vect{k})$ with $\vect{k}=(0,\pi)$ have the opposite sign, 
so that the gap equation in (\ref{gap eq}) is satisfied with 
$V_{{\rm sin}}(\pi,\pi)>0$.  

For Sr$_2$RuO$_4$ we consider that a kind of {\it nesting} property 
of the Fermi surface plays an important role.  
This is the second point of our mechanism.  
Figure 1 shows a schematic Fermi surface for the $\gamma$ band.  
Since the AF fluctuation in Sr$_2$RuO$_4$ has momentum
$\vect{q}_0$, the Fermi surface is also shifted by 
$(2\pi/3, 2\pi/3)$ in Fig.\ 1.  
It is apparent that some part of the shifted Fermi surface overlaps with 
the original Fermi surface with modulo $2\pi$.  
In analogy to the case of high-$T_{\rm c}$ superconductivity, 
if the superconducting
order parameters have the  opposite sign on these overlapping portions 
of the Fermi surface, the gap equation is satisfied with 
$V_{\rm tri}^{(a)}(2\pi/3,2\pi/3)>0$.  
From Fig.\ 1, it is natural to consider the p-wave pairing 
instead of the singlet d$_{x^2-y^2}$-wave pairing.  
\begin{figure}
\psfig{figure=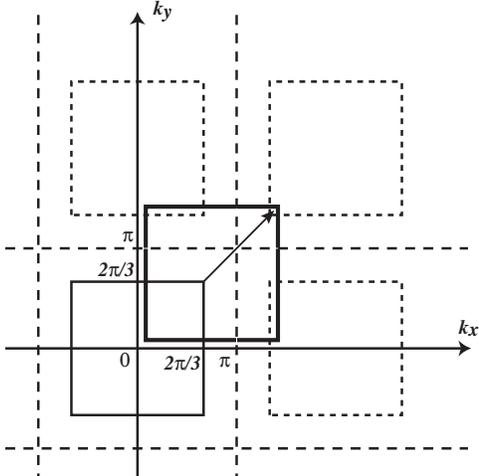,height=6.5cm}
\caption{A schematic Fermi surface for the $\gamma$ band of 
Sr$_2$RuO$_4$.  In order to show the nesting property, the Fermi surface 
shifted by the AF wave number, ${\mib q}_0$, is also shown by the solid line.  
The thin dashed lines indicates the Fermi surfaces in the extended 
Brillouin zone.}
\end{figure}

In order to clarify this point quantitatively, we compare various 
kinds of anisotropic superconductivity using the effective 
interaction and the simplified Fermi surface.  
Near the transition temperature $T_{\rm c}$, we rewrite the gap 
equations as
\begin{equation}
\phi(\vect{k})=-\sum_{k'}V_{\phi}(\vect{k}-\vect{k}')
\phi(\vect{k}')\frac{1}{2\xi_{k'}}
\tanh \frac{\beta_c \xi_{k'}}{2},
\label{gapeq2}
\end{equation}
where $\phi(\vect{k})$ represents one of the order parameters 
$\Delta(\vect{k})$ or $d^{(a)}(\vect{k})$, and $V_{\phi}$ is determined 
from Eqs.\ (\ref{Vsin}) depending on $\phi$. 
In the weak coupling approximation, $T_{\rm c}$ is obtained as
\begin{equation}
k_B T_{\rm c}=1.13 \hbar v_F k_c \exp \left[-\frac{1}{N(0) 
\langle \langle V_\phi \rangle \rangle_{FS}} \right],
\end{equation}
where $v_F$, $k_c$ and $N(0)$ are the Fermi velocity, cut-off of the 
wave number, and the density of states at the Fermi energy, respectively.  
$\langle \langle V_\phi \rangle \rangle_{FS}$ means the 
average over the Fermi surface, 
\begin{equation}
\langle \langle V_\phi \rangle \rangle_{FS} \equiv 
-\frac{\int_{FS}d\vect{k}\int_{FS}d\vect{k}'V_{\phi}(\vect{k}-\vect{k}')
\phi(\vect{k})\phi(\vect{k}')}
{\left[ \int_{FS}d\vect{k}' \right] \int_{FS}d\vect{k}\phi^2(\vect{k})}.
\label{vphi}
\end{equation}
We identify that the order parameter which gives the largest 
$N(0)\langle \langle V_\phi \rangle \rangle_{FS}$ is realized.  

For Sr$_2$RuO$_4$ 
we choose order parameters $\phi(\vect{k})$ as follows
\begin{eqnarray}
\phi_1(\vect{k})&=&\cos{k_x}+\cos{k_y},\nonumber\\ 
\phi_2(\vect{k})&=&\cos{k_x}-\cos{k_y},\nonumber\\
\phi_3(\vect{k})&=&\sin{k_x}\sin{k_y},\nonumber\\ 
\phi_4(\vect{k})&=&\sin{k_x}, (\hat{\vect{d}} \perp \hat{\vect{z}}),\nonumber\\
\phi_5(\vect{k})&=&\sin{k_x}, (\hat{\vect{d}} \parallel \hat{\vect{z}}),
\end{eqnarray}
where $\phi_1 \sim \phi_3$ correspond to singlet pairings, and 
$\phi_4$, $\phi_5$ to triplet pairings, respectively.  
The most probable candidate for Sr$_2$RuO$_4$ is 
$\hat{\vect{z}}(\sin{k_x}\pm i\sin{k_y})$ which is equivalent to 
$\phi_5$ just below $T_{\rm c}$, 
because the gap equation (\ref{gapeq2}) for $\sin{k_x}\pm i\sin{k_y}$ 
is exactly same as that for $\phi_5$.  
If $N(0)\langle \langle V_{\phi_5} \rangle \rangle_{FS}$ is the largest,
we expect that the 
order parameter $\vect{d}(\vect{k})=\hat{\vect{z}}(\sin{k_x}\pm i\sin{k_y})$
is realized, because near zero temperature 
it acquires a larger energy gap than $\phi_5$.  

To emphasize the characteristic feature of the nesting, 
we assume the simplified Fermi surface as shown in Fig.\ 1.  
For the $\vect{q}$ dependence of $\chi_{(\uparrow \uparrow,\ 0)}(\vect{q})$
with a maximum at ${\mib q}_0$, 
we use the susceptibility obtained in the LDA calculation\cite{mazin1}, 
and fix $S(\vect{0})=0.8$ with 
$S(\vect{q})\equiv\frac{I}{N}\chi_{(\uparrow\uparrow,\ 0)}(\vect{q})$.  
We regard $S(\vect{q}_0)$ as a phenomenological parameter.  

Figure 2 shows the $\alpha$ dependence of 
$N(0)\langle \langle V_{\phi_n} \rangle \rangle_{FS}$ $(n = 1\sim 5)$ 
for $S(\vect{q}_0)=0.95$.  
We examined various choices of $S(\vect{q}_0)$ from $0.90$ to $0.99$ to 
find that the results do not change qualitatively.  
When the anisotropy is weak ($\alpha \sim 1$), the singlet 
d$_{x^2-y^2}$-wave superconductivity, $\phi_2$, is stabilized.  
On the other hand, when $\alpha$ is small, the order parameter $\phi_5$ 
is stabilized which is consistent with experiments.  
\begin{figure}
\psfig{figure=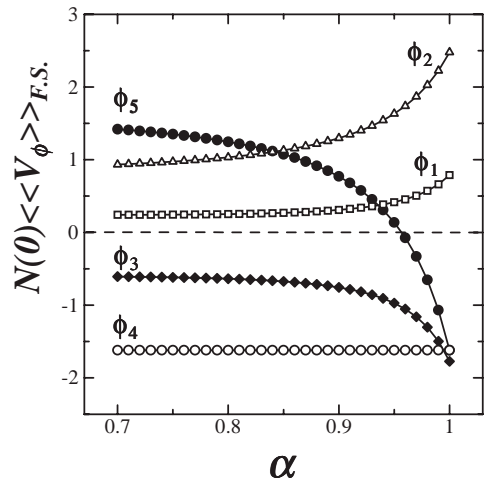,height=6.5cm}
\caption{The dependence of the anisotropy parameter, $\alpha$, of 
$N(0)\langle \langle V_{\phi_n} \rangle \rangle_{FS}$ $(n=1\sim5)$ 
for $S({\mib q}_0)=0.95$.}
\end{figure}

The phase diagram as a function of $\alpha$ and $S(\vect{q}_0)$ is 
determined by examining various values of $S(\vect{q}_0)$. 
Because it is unphysical to assume that $S(\vect{q}_0)$ is very 
close to 1, we show the results up to $S(\vect{q}_0)= 0.99$ in Fig.\ 3. 
When the spin fluctuation is isotropic (i.e., $\alpha=1$), the singlet 
d$_{x^2-y^2}$-wave superconductivity, $\cos{k_x}-\cos{k_y}$, 
is the most stable pairing.  
This is consistent with the previous studies\cite{mazin1}.
However we find a fairly large parameter region where 
the state corresponding to $\hat{\vect{z}}(\sin{k_x}\pm i\sin{k_y})$ 
is realized.  
\begin{figure}
\psfig{figure=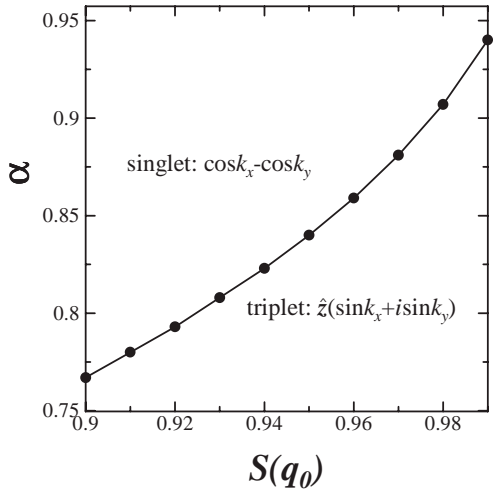,height=6.5cm}
\caption{Phase diagram as a function of the anisotropy parameter, $\alpha$, 
and $S({\mib q}_0)$ which is the maximum of 
$S({\mib q})=(I/N)\chi_{(\uparrow\uparrow,\ 0)}({\mib q})$, with 
${{\mib q}_0}=(\pm 2\pi/3,\pm 2\pi/3)$.}
\end{figure}

Finally we discuss the competition between the singlet $\cos{k_x}-\cos{k_y}$ 
pairing and the triplet $\hat{\vect{z}}(\sin{k_x}\pm i\sin{k_y})$ pairing 
in terms of the effective interaction and the nesting property.  
From the explicit form of $V_{\phi_n}$ for $n=2$ and 5, 
we can see that a relation $V_{\phi_{2}}\geq V_{\phi_{5}}$ is satisfied.  
Therefore if we consider only the magnitude of the effective interaction, 
the singlet pairing is favorable.  
However the nesting property favors the triplet pairing.  
Let us assume that $V_{\phi_{n}}$ is enhanced very strongly by 
the AF fluctuation and approximated as
\begin{equation}
V_{\phi_{n}}(\vect{q})=\frac{I}{N} 
A_n \hat{\delta}(q_x \pm 2\pi/3) \hat{\delta}(q_y \pm 2\pi/3),
\end{equation}
with $\hat{\delta}$ being the $\delta$ function with modulo $2\pi$.  
Using this approximated form of $V_{\phi_{n}}(n=2\ {\rm and}\ 5)$, 
we obtain
\begin{eqnarray*}
N(0)\langle \langle V_{\phi_2} \rangle \rangle_{FS} &=& 
   [-2.79\times 10^{-2}\delta(0)+4.91\times 10^{-2}] A_2, \nonumber\\
N(0)\langle \langle V_{\phi_5} \rangle \rangle_{FS} &=& 
   [4.24\times 10^{-2}\delta(0)+5.06\times 10^{-2}] A_5.
\end{eqnarray*}
This estimation shows that the $\hat{\vect{z}}(\sin{k_x}\pm i\sin{k_y})$ 
pairing utilizes the peak of $\chi_{(\uparrow\uparrow,\ 0)}(\vect{q})$ 
at $\vect{q}_0$ more effectively than $\cos{k_x}-\cos{k_y}$ pairing does.  
Therefore, even if $A_2 > A_5$, the triplet pair can be stabilized.  

In determining the phase diagram in Fig.\ 3, we have assumed simple 
functional forms of the order parameters, $\phi_n({\vect{k}})$.
For the detailed calculations, it will be necessary to optimize the 
$\vect{k}$-dependence of $\phi_n({\vect{k}})$.  
However the global feature of the phase diagram will not change.  

In summary, we have generalized the RPA formulation of the effective 
interaction due to the spin fluctuations 
and derived gap equations including the anisotropic case.  
We have shown that the state corresponding to 
$\hat{\vect{z}}(\sin{k_x}\pm i\sin{k_y})$ becomes the most stable 
even if the AF spin fluctuation is dominant, when the anisotropy is 
strong enough  
and the nesting property of the Fermi surface is present.  
Although the nesting for the actual Fermi surface will be 
weaker than what we assumed here, it is reasonable to think 
that our mechanism is the most promising one as far as the AF 
fluctuation is dominant.  

In this paper we have investigated the pairing in the $\gamma$ band.  
However it is straightforward to consider the other bands 
($\alpha$ and $\beta$ bands) in Sr$_2$RuO$_4$.  
Since the nesting property will be comparable or even stronger 
for these bands than for the $\gamma$ band, we expect the same 
mechanism for triplet superconductivity for $\alpha$ and $\beta$ 
bands even if the $\gamma$ band does not have the peak near $\vect{q}_0$.  

It is reported that Sr$_2$RuO$_4$ has exotic property called 
as {\it 3K phase}\cite{maeno2} when Ru metal is embedded in the 
single crystal.  
We speculate that the enhancement of $T_{\rm c}$ is due to the 
increase of the anisotropy (i.e., decrease of $\alpha$) 
near the interface region between Sr$_2$RuO$_4$ and Ru metal.  
We consider that to investigate the origin of the anisotropy is very 
important both for understanding the superconductivity and for finding 
the new exotic phenomena.  

We are grateful for useful discussions with M.\ Sigrist, Y.\ Maeno and 
Y.\ Matsuda.  
One of the authors (T.K.) thanks to H.\ Namaizawa for useful instructions.


%
%

%
%
\end{document}